\documentclass[12pt]{article}
\pdfoutput=1
\usepackage[lmargin=0.75in,rmargin=0.75in,tmargin=1.0in,bmargin=1.0in]{geometry}

\usepackage{wrapfig}
\usepackage{graphicx}
\usepackage{tabularx}
\usepackage{pdfpages}
\usepackage{float}
\usepackage[font={small}]{caption} 
\usepackage[square,numbers,sort]{natbib}
\usepackage{authblk}

\usepackage{listings}
\usepackage[hyphens]{url}
\usepackage[hidelinks]{hyperref}
\begin{document}

\date{}

\title{Structured dataset of reported cloud seeding activities in the United States (2000–2025) using an LLM} 

\author{
\begin{minipage}[t]{0.45\textwidth}
  \centering
  \textbf{Jared Joseph Donohue}\textsuperscript{*}\\
  \small \textit{*Corresponding author}\\
  \small Data Science Institute\\
  \small Columbia University\\
  \small New York, NY, USA\\
  \small jjd2203@columbia.edu\\
\end{minipage}
\hfill
\begin{minipage}[t]{0.45\textwidth}
  \centering
  \textbf{Kara D. Lamb}\\
  \small Department of Earth and\\ \small Environmental Engineering\\
  \small Columbia University\\
  \small New York, NY, USA\\
  \small kl3231@columbia.edu
\end{minipage}
}

\maketitle

\section*{Abstract} 
\textbf{Cloud seeding, a weather modification technique used to increase precipitation, has been practiced in the western United States since the 1940s. However, comprehensive datasets are not currently available to analyze these efforts. To address this gap, we present a structured dataset of reported cloud seeding activities in the U.S. from 2000–2025, including the project name, year, season, state, operator, seeding agent, apparatus used for deployment, stated purpose, target area, control area, start date, and end date. Combining our multi-stage PDF-to-text extraction pipeline with OpenAI's \textit{o3} large language model (LLM), we processed 832 historical reports from the National Oceanic and Atmospheric Administration (NOAA). The resulting dataset demonstrates 98.38\% estimated accuracy, based on manual review of 200 randomly sampled records, and is publicly available on Zenodo. This dataset addresses the gap in cloud seeding data and demonstrates the potential for LLMs to extract structured information from historical environmental documents. More broadly, this work provides a scalable framework for unlocking historical data from scanned documents across scientific domains.}

\section*{Background \& Summary}
Cloud seeding is a weather modification technique used to enhance precipitation, typically in regions experiencing drought, water scarcity, or to support snowpack accumulation. In the United States, cloud seeding experiments began in the 1940s \cite{dri2023cloudseeding}, and systematic reporting was introduced with the Weather Modification Reporting Act of 1972 \cite{cfr908}. The Act requires individuals and organizations to notify the U.S. Department of Commerce at least 10 days before and after conducting weather modification activities, using Form 17-4 (Initial Report on Weather Modification Activities) and Form 17-4A (Interim Activity Reports and Final Report). These forms are intended to be submitted to NOAA and archived on the NOAA Weather Modification website \cite{noaa_wm_reports}. However, only reports from 2000 to 2025 are currently available online. Records from earlier years are not accessible through NOAA’s website and have not yet been located elsewhere. As a result, our dataset covers only the 2000–2025 period, representing the full set of publicly accessible reports to date.

While the NOAA reports are made publicly available online, they are stored individually as scanned PDFs with inconsistent formatting and structure. This lack of standardization makes it difficult to extract key information at scale, or integrate the reports into other structured datasets. As a result, data on cloud seeding in the U.S., such as locations, dates, purposes, cloud seeding agent, and agent deployment methods, remains largely inaccessible for analysis.

To solve this data access problem, we created a structured dataset from 832 NOAA cloud seeding reports spanning from 2000 to 2025. Our work transforms these scattered, inconsistently formatted documents into a single, structured CSV file that enables the analysis of U.S. weather modification activities over time. The dataset achieves an estimated 98.38\% accuracy across all extracted fields, based on a manual evaluation of 200 randomly sampled records (out of the total population of 832), each reviewed by two independent human annotators against the original PDFs.

This data extraction method can be extended to similar government-mandated environmental reporting systems, such as those related to water usage, air quality monitoring, or land management, many of which exist in PDF form with highly variable formatting. By transforming previously inaccessible records into structured data, this work lays the groundwork for evidence-based research and policy evaluation in an area of growing environmental interest\cite{gaoereport}.

Beyond its methodological contributions, the resulting dataset has potential value across a wide range of applications. Researchers can use it to study long-term patterns in weather modification practices, analyze the use and evolution of different seeding agents and deployment methods, or assess geographic and seasonal trends in operational activity.

For example, while there have been several recent research campaigns focused on process-level understanding and evaluation of cloud seeding, such as the the 2012 ASCII (AgI Seeding Cloud Impact Investigation) over the Sierra Madre range in Wyoming, or the 2017 Seeded and Natural Orographic Wintertime clouds: the Idaho Experiment (SNOWIE) campaign in Western Idaho\cite{geerts2013agi, tessendorf_transformational_2019}, there has been little quantitative analysis of reported cloud seeding activities. With recent renewed interest in both traditional weather modification technology (e.g. the UAE's Rain Enhancement Program and several recent commercial start-ups), as well as geoengineering strategies focused on Solar Radiation Management such as Marine Cloud Brightening that use similar intervention strategies to mitigate climate change\cite{feingold2024physical}, this dataset can provide historical context for weather modification usage in the United States. In addition, this dataset can provide insights into current limitations in reporting requirements that impede large-scale evaluation of the environmental and meteorological implications of weather modification\cite{gaoereport}. 

\section*{Methods}

\begin{figure*}[htbp]
    \centering
    \includegraphics[width=0.9\linewidth]{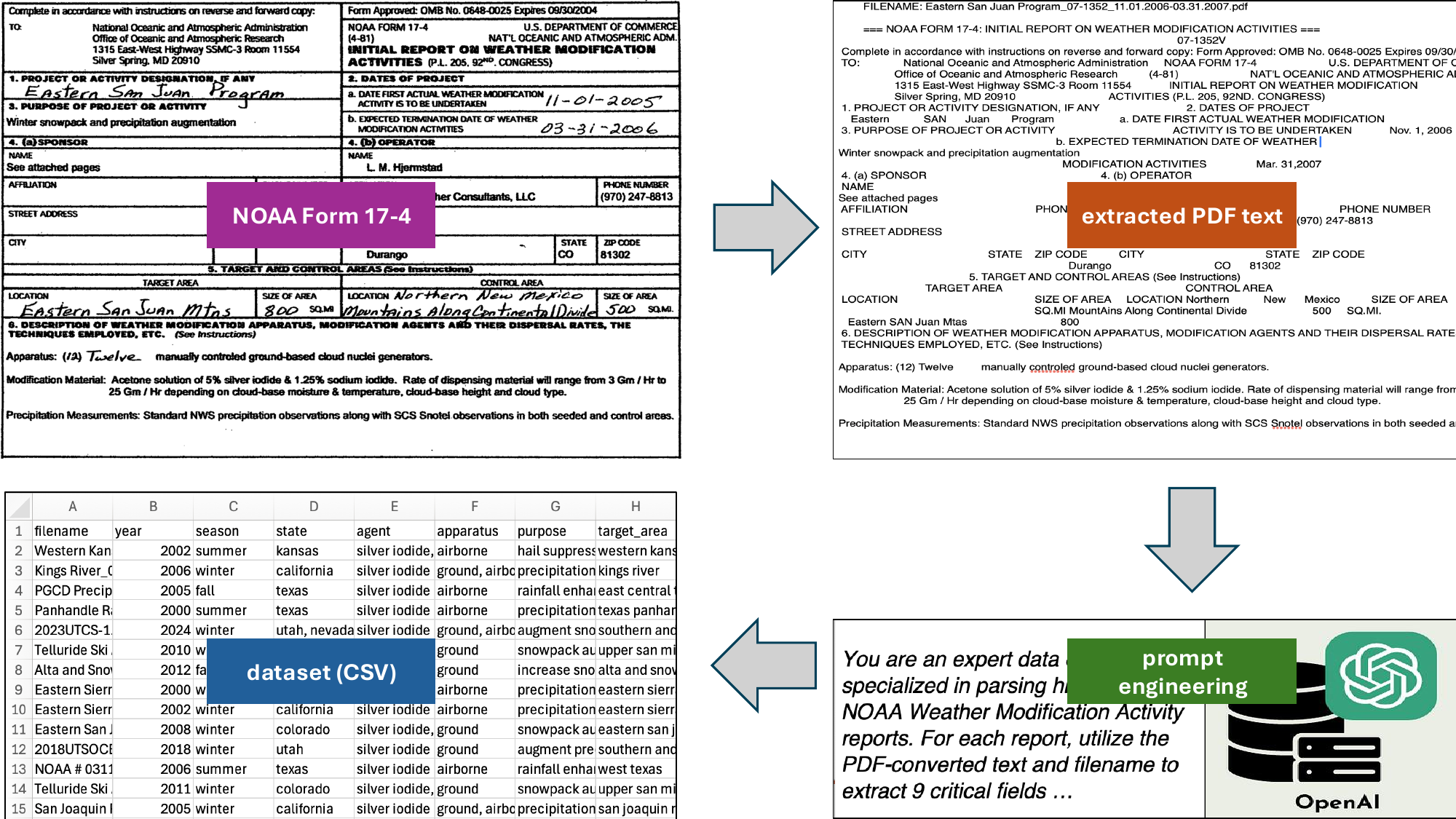}
    \caption{NOAA Form~17-4 Dataset Extraction Pipeline}
    \label{fig:system_diagram}
\end{figure*}

\subsection*{Input Data Source}

Cloud seeding reports were manually downloaded from the NOAA Central Library’s Weather Modification Project Reports Archive (\url{https://library.noaa.gov/weather-climate/weather-modification-project-reports}) \cite{noaa_wm_reports}. A total of 1,025 PDF files were collected, including Form~17-4, Form~17-4A, and other supplemental documents. These 1,025 files represent all publicly reported weather modification activities submitted to NOAA. Since the reports are regulatory submissions intended for public access, they are in the public domain and their reuse and redistribution are permitted under U.S. government policy.

Each PDF included Form~17-4, Form~17-4A, or both, along with supplemental pages. Form~17-4 contains the main project summary, including the project purpose, operational period, operator, seeding agents, apparatus for deployment, and target/control areas. Form~17-4A provides tabular data on seeding activity, including days seeded per month and agent quantities released.

For our dataset, we focused exclusively on extracting fields from Form~17-4 which contained the majority of relevant project descriptive information. While many PDFs included supporting materials, such as maps, meteorological analyses, operational logs, and narrative summaries, these were excluded due to inconsistent formatting that would require extensive preprocessing with minimal dataset enhancement.

\subsection*{Preprocessing}
A custom Python pipeline was developed to preprocess all Form~17-4 PDF documents (Figure~\ref{fig:system_diagram}). First, all downloaded files were stored in a local directory. In many cases, a single cloud seeding project was documented across multiple submissions, indicated by suffixes such as \textit{.M} (initial report), \textit{.I} (interim report), and \textit{.F} (final report). These reports were consolidated into a single file per project using a combination of programmatic and manual methods to preserve project-level information. After consolidation, 832 unique weather modification projects remained for processing and inclusion in the dataset.

The scanned documents included both typed and hand-written text, and sometimes a mix of both in the same document. To convert the PDFs into plain text for further processing, a combination of three technologies was used. For native, text-based documents (57\% of documents), text was extracted using the \texttt{pymupdf} Python library \cite{pymupdf4llm}. For scanned or hand-written documents (43\% of documents), optical character recognition (OCR) was applied using a combination of \texttt{pytesseract}, an open-source wrapper for the Tesseract OCR engine \cite{tesseract2007} (processed 16\% of all documents), and \texttt{llm-whisperer} (processed 27\% of all documents), a layout-preserving PDF extraction technology that combines OCR and native-text extraction \cite{llmwhisperer_python_client}. All code for the preprocessing pipeline, including file merging, text extraction, and LLM-integration, is publicly available on GitHub (\url{https://github.com/jdonohue44/NOAA-Weather-Modification-Forms-LLM-Extractor}) \cite{llm_extractor_github}.

In addition to the extracted text, filenames were also provided to the model because the filenames often contained useful contextual clues about the report's location or time frame (e.g. \textit{North Dakota Cloud Modification Project - District I\_02-1147\_06.01.2002-08.31.2002.pdf}). Notably, there appeared to be two distinct file naming conventions in the archive. Older files (generally before 2017) tend to use a descriptive format that includes the location and date (e.g., \textit{Colorado-SanJuan-2003.pdf}), whereas newer files (from around 2018 onward) follow a standardized alphanumeric format such as \textit{2018TXSWMA-1\_2018.pdf}, which embeds the year, state abbreviation, and project code. These filenames were leveraged during extraction to assist with metadata recovery.

\subsection*{LLM Integration}
We integrated with OpenAI’s platform of LLMs to analyze the extracted text and synthesize the key metadata fields for the structured dataset. After evaluating multiple models, the \textit{o3} model demonstrated the best performance and was selected for all extractions \cite{openai_models}.

\subsection*{LLM Prompt}
We experimented with multiple prompt templates using prompt engineering techniques to improve extraction quality. The final prompt, shown below with truncation, employs chain-of-thought reasoning by guiding the model through intermediate steps before producing structured outputs. This reasoning style improves performance on tasks requiring multi-step inference and disambiguation \cite{wei2022chainofthought}. The prompt was used with the \textit{o3} reasoning model for all extractions in the final dataset. All prompt versions and evaluation scripts are available in our public GitHub repository \cite{llm_extractor_github}.

\vspace{1em}

\begin{lstlisting}[breaklines=true]
# NOAA Weather Modification Report Extraction Expert
You are an expert data extractor specialized in parsing historical NOAA Weather Modification Activity reports. For each report, utilize the PDF-converted text and filename to extract **12** critical fields.

## Instructions
For each field, carefully analyze all available information using a step-by-step reasoning process. Reason step-by-step internally, using evidence from the filename and report content to resolve conflicting or ambiguous information. Output only the final extracted fields using the format and rules provided below.

## Fields to Extract
[truncated for brevity]

## Example Reasoning (Internal)
[truncated for brevity]

## Final Extracted Fields Format
Present your final extracted results concisely as follows, in lowercase, comma-separating multiple values when applicable. Do not include commentary, explanations, or placeholders. Leave field blank if truly unknowable after exhausting all inference methods using the filename and text evidence.

PROJECT: [extracted value]
YEAR: [extracted value]  
SEASON: [extracted value]  
STATE: [extracted value]  
OPERATOR AFFILIATION: [extracted value]  
AGENT: [extracted value]  
APPARATUS: [extracted value]  
PURPOSE: [extracted value]  
TARGET AREA: [extracted value]  
CONTROL AREA: [extracted value]  
START DATE: [extracted value]  
END DATE: [extracted value]
\end{lstlisting}

\subsection*{Postprocessing}
As instructed, the LLM analyzed the extracted PDF text and responded with a list containing the key information we requested. In order to neatly store this information into CSV format, we then mapped key values into more concise column names, stripped any blank space, and converted to lowercase. Using the formatted data object, each project could be easily written to CSV, with one row per object and columns corresponding to the extracted fields. Lastly, to prepare the final dataset, we used \texttt{clean-dataset.py} in \cite{llm_extractor_github} to remove any duplicates, standardize formatting, and fill blank cells with \texttt{NaN}.



\section*{Data Records}
\begin{figure}[H]
    \centering
    \includegraphics[width=0.9\linewidth]{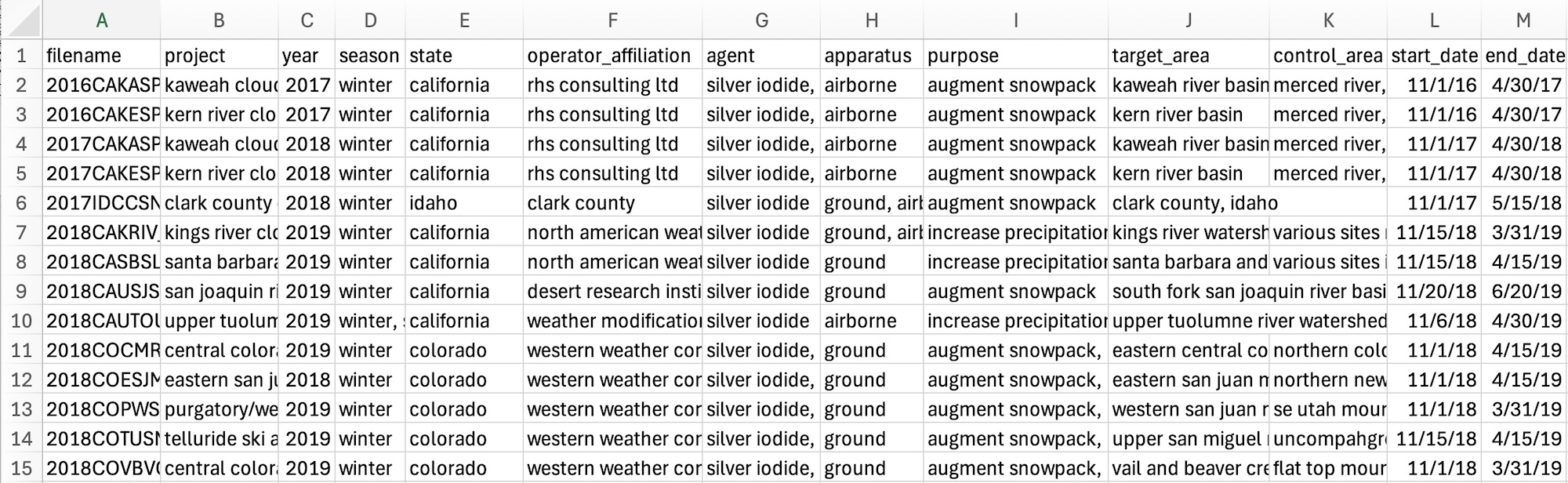}
    \caption{Screenshot of the CSV Dataset (\textit{cloud\_seeding\_us\_2000\_2025.csv})}
    \label{fig:dataset_screenshot}
\end{figure}

The dataset is available on Zenodo as a single CSV file (\url{https://doi.org/10.5281/zenodo.14925811}) \cite{final_dataset}. It contains structured information from 832 unique cloud seeding project reports across the United States from 2000 to 2025, with one row per project and one column per extracted field (Figure~\ref{fig:dataset_screenshot}).

The following fields are present in the dataset:
\begin{itemize}
\item \textbf{filename}: The original name of the file from which the data was extracted.
\item \textbf{project}: The official name or designation of the cloud seeding project or activity.
\item \textbf{year}: The calendar year in which the cloud seeding activity occurred (e.g., 2015).
\item \textbf{season}: The meteorological or operational season during which the activity took place (e.g., winter, summer).
\item \textbf{state}: The U.S. state where the cloud seeding operation was conducted (e.g., California).
\item \textbf{operator\_affiliation}: The organization or agency responsible for conducting or overseeing the seeding operation.
\item \textbf{agent}: The material used to seed clouds (e.g., silver iodide, dry ice).
\item \textbf{apparatus}: The method or equipment used to deliver the seeding agent (e.g., ground-based generators, aircraft).
\item \textbf{purpose}: The stated objective of the cloud seeding activity (e.g., augment snowpack, increase rainfall, suppress hail).
\item \textbf{target\_area}: The geographical area where the cloud seeding was intended to take effect.
\item \textbf{control\_area}: A designated area used as a comparison region, typically not subject to cloud seeding.
\item \textbf{start\_date}: The first reported date of seeding activity for the project.
\item \textbf{end\_date}: The final reported date of seeding activity for the project.
\end{itemize}

\subsection*{Data Visualizations}
\begin{figure}[H]
\centering
\includegraphics[width=0.9\linewidth]{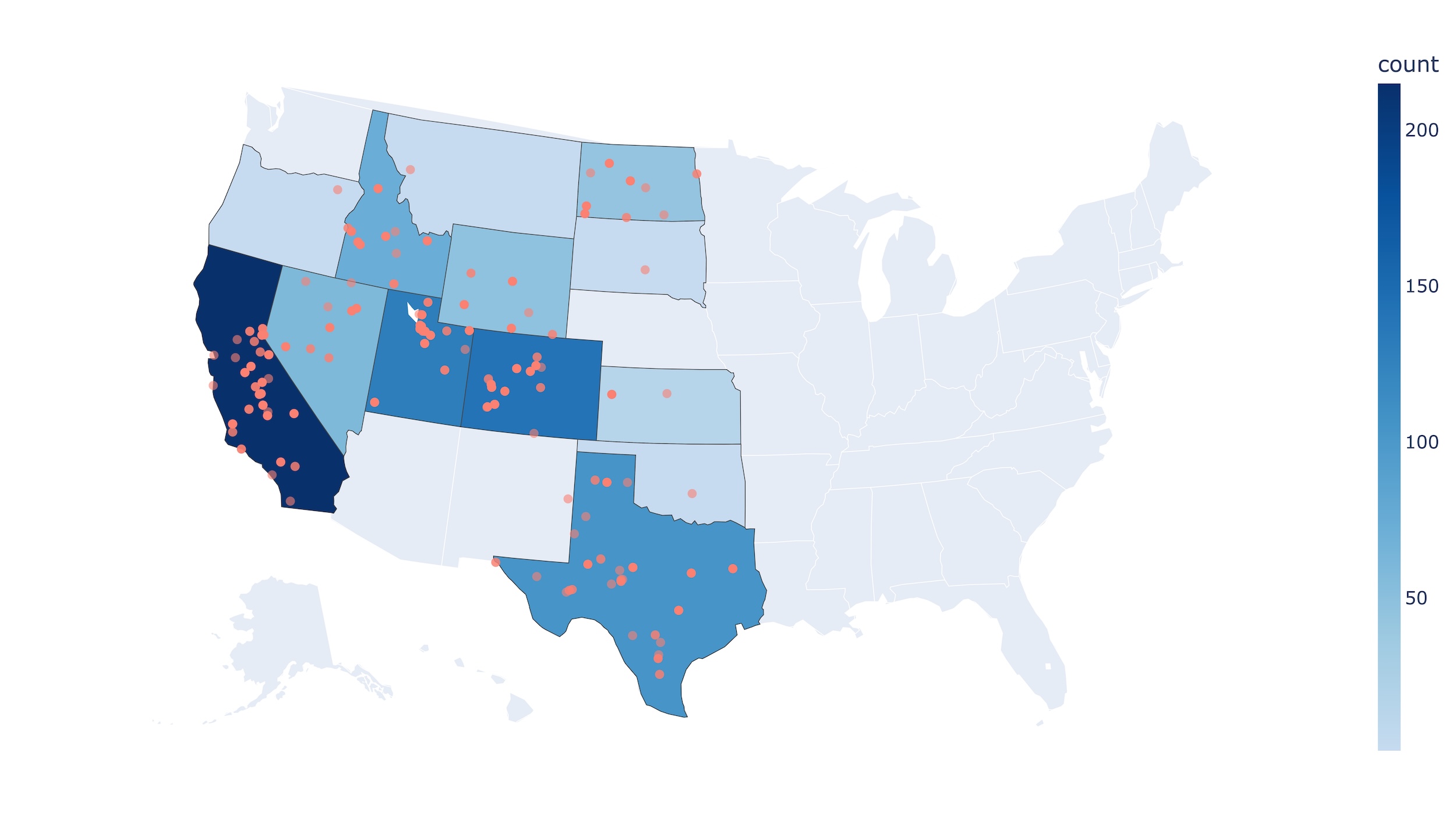}
\caption{Cloud Seeding Activity by U.S. State (2000–2025). States with active weather modification programs show the highest number of recorded operations. Specific locations (shown as salmon markers) are from geocoding the stated \textit{target\_area} field with the GoogleMaps API.}
\label{fig:cloud_seeding_map_locations}
\end{figure}

Between 2000 and 2025, cloud seeding activity in the United States was geographically concentrated in western states that rely on snowpack for water supply. California, Colorado, and Utah accounted for the majority of seeding activities, with summertime rain enhancement in Texas also contributing a substantial number of activities (Figure~\ref{fig:cloud_seeding_map_locations}). These records align with states that maintain active cloud seeding programs \cite{gaoereport}.

\begin{figure}[H]
\centering
\includegraphics[width=0.9\linewidth]{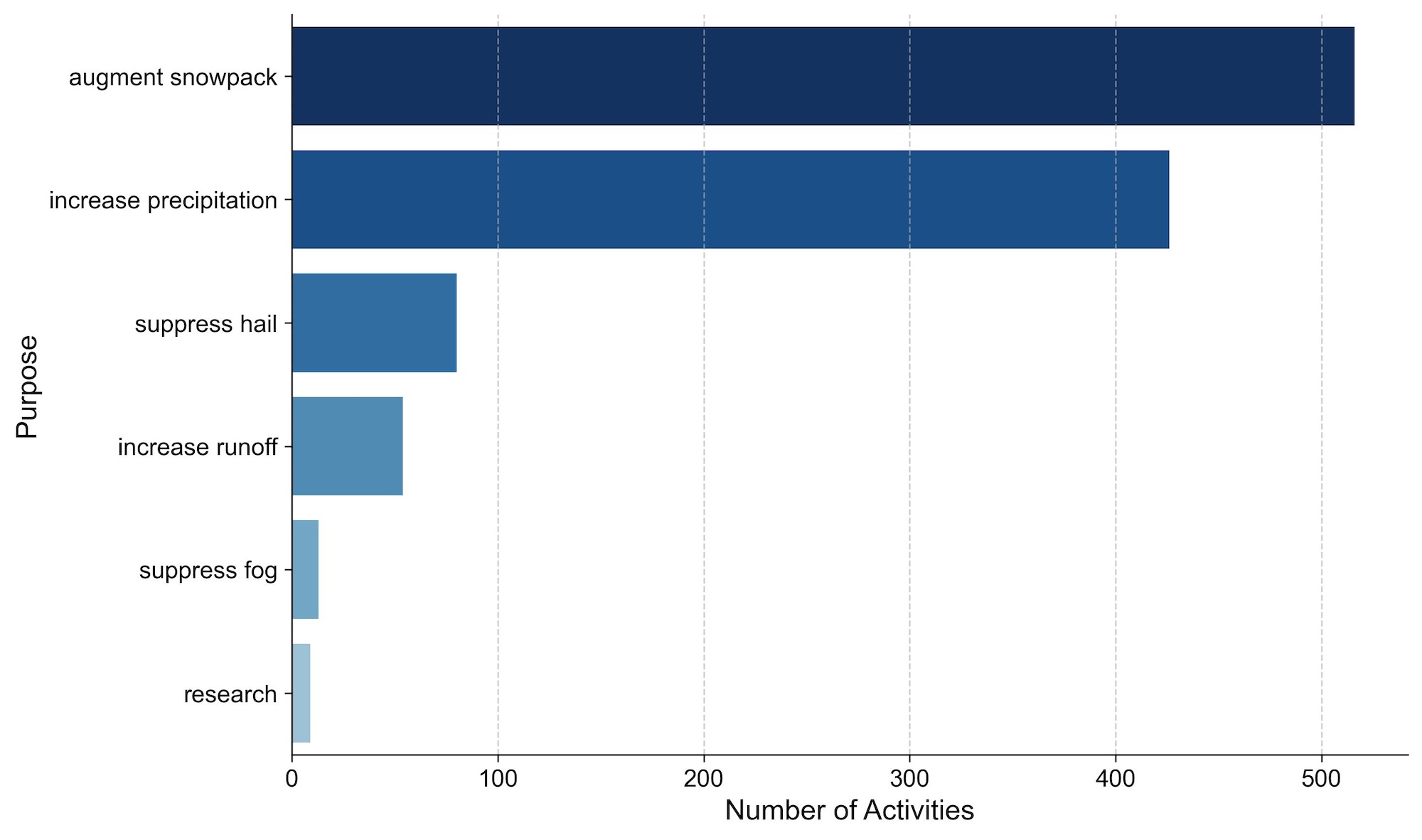}
\caption{Stated Purpose of Cloud Seeding Activity (2000–2025). Augmenting snowpack is the leading purpose, followed by increasing precipitation.}
\label{fig:cloud_seeding_purpose}
\end{figure}

The primary stated purpose of cloud seeding events was to increase snowpack, followed by increasing precipitation and rain (Figure~\ref{fig:cloud_seeding_purpose}). Snowpack enhancement is particularly important in mountainous regions where runoff from melting snow supplies reservoirs and irrigation systems.

\begin{figure}[H]
\centering
\includegraphics[width=0.9\linewidth]{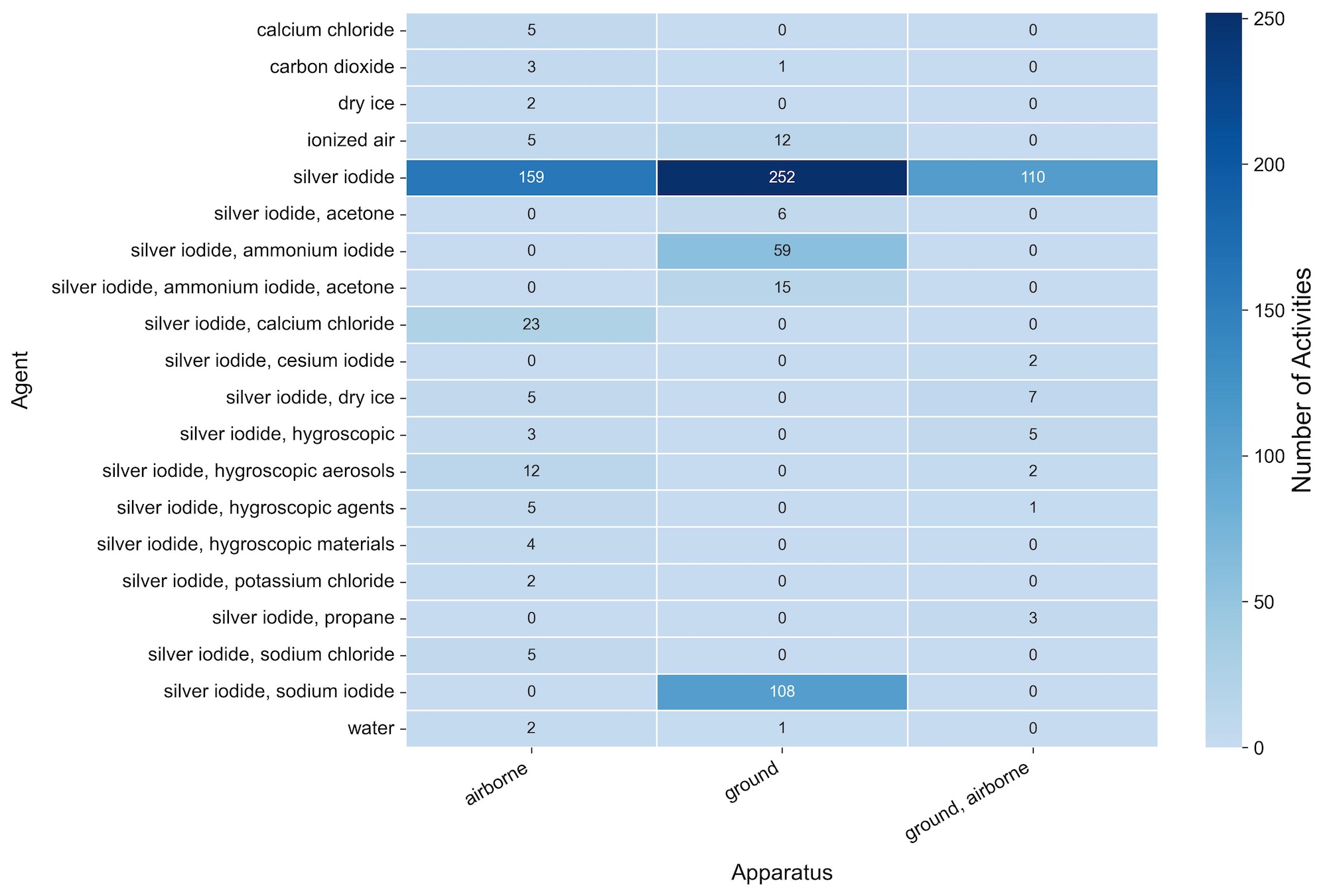}
\caption{Agent and Apparatus Used for Cloud Seeding Operations (2000–2025). Silver iodide dominates among agents and is deployed most frequently using a ground-based apparatus.}
\label{fig:cloud_seeding_agent_and_apparatus}
\end{figure}

Silver iodide is by far the most common seeding agent, especially in ground-based and airborne operations (Figure~\ref{fig:cloud_seeding_agent_and_apparatus}). Other agents such as calcium chloride and carbon dioxide are used less frequently. Ground deployment is the most prevalent deployment method, particularly when using silver iodide, although a mix of ground and airborne (ground, airborne) approaches is also present.

\begin{figure}[H]
\centering
\includegraphics[width=0.9\linewidth]{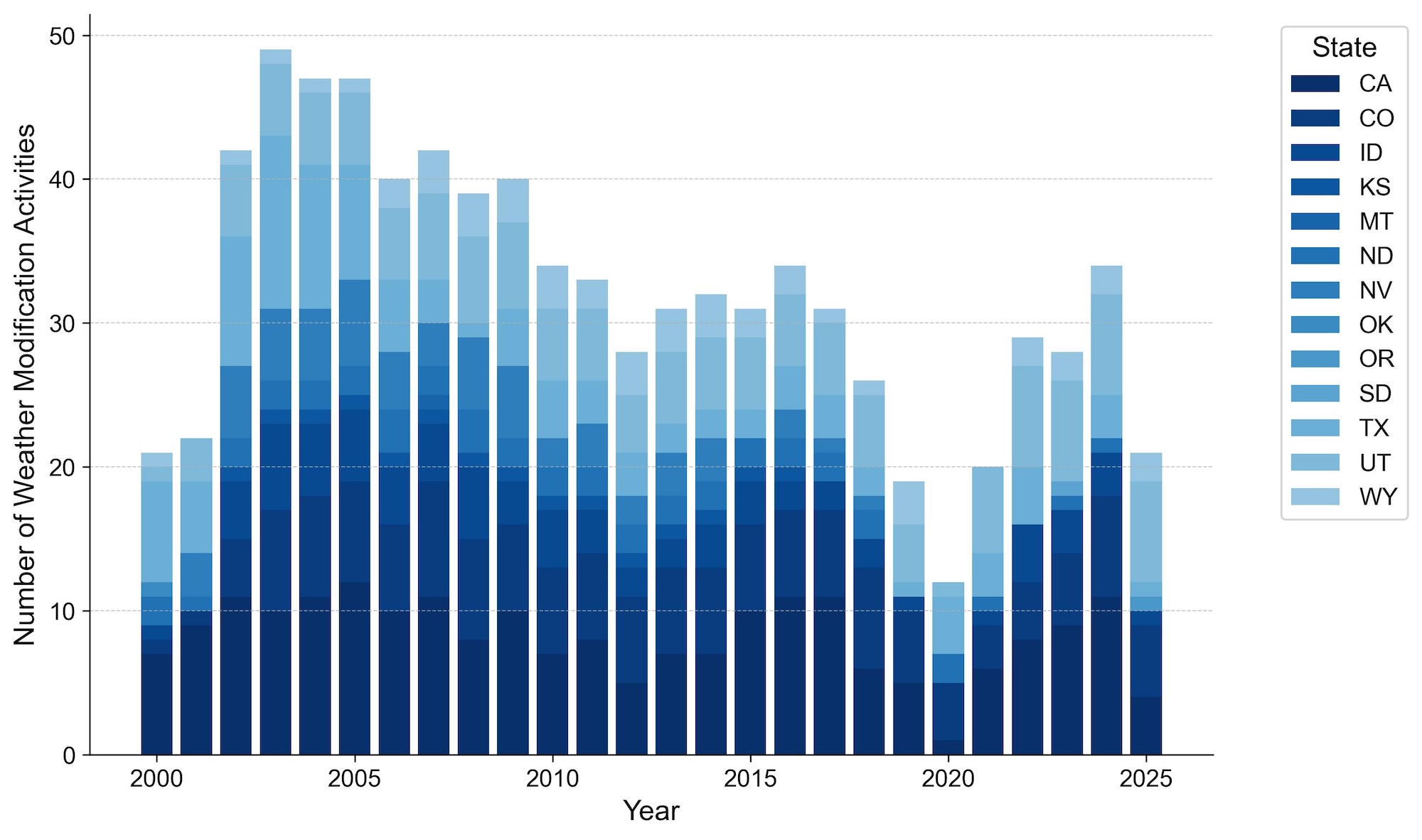}
\caption{Cloud Seeding Activity by U.S. State over Time (2000–2025). The number of activities peaked between 2003-2005, declined gradually, and rose again after 2021.}
\label{fig:cloud_seeding_timeseries}
\end{figure}

Over this 25 year period, the number of weather modification events peaked in the early to mid-2000s, then declined through the 2010s, before rebounding in 2024 and 2025 (Figure~\ref{fig:cloud_seeding_timeseries}).

\section*{Technical Validation}
To address the potential for LLMs to hallucinate content not present in the source documents, and the inconsistent quality of the input PDFs themselves, we measured field-level extraction accuracy by manually reviewing a random sample of 200 records extracted by the LLM and comparing them to ground truth. This evaluation provides an empirical bound on potential extraction errors in the released dataset, enabling downstream users to incorporate measurement uncertainty into their analyses.

Although LLMs are capable of producing fluent and structured text, they can generate information that is not present in the source material \cite{huang2025survey}. In the context of this paper, such hallucinated content could include incorrect dates, locations, or project descriptions, which would compromise the reliability of downstream analyses involving historical weather modification activities in the United States.

Additionally, the quality of the input documents presented another challenge. Many PDFs contained handwritten annotations, low-resolution scans, or incomplete text referring the reader to information from a previous year or external attachments rather than providing complete answers in the designated fields. For example, reports from the North Dakota Cloud Modification Project and the Western Kansas Weather Modification Program left fields blank and instead referred to separate documents such as an operational plan or an attached sheet. These references could not be processed directly, limiting the ability of the model to extract a full set of structured values.

\subsection*{Extraction Accuracy}
From the population of N=832 records we randomly sampled n=200 to evaluate field-level extraction accuracy (Table~\ref{tab:field-accuracy-evaluations}). This sample size was chosen to yield a margin of error of $\pm 3$ percentage points at the 95\% confidence level. The evaluation was conducted by two independent human experts manually creating a golden dataset (ground truth) using the source PDFs, then comparing the extracted fields against the golden dataset using \texttt{compare-to-golden.py} \cite{llm_extractor_github}.

\begin{table}[H]
\centering
\caption{Field-Level Extraction Accuracy (n=200; model=\textit{o3})}
\begin{minipage}{0.85\textwidth}
\begin{tabularx}{\textwidth}{|X|c|c|}
\hline
\textbf{Field} & \textbf{Accuracy (\%)} & \textbf{95\% Confidence Interval} \\
\hline
Project & 100.00\% & [97.00\%, 100.00\%]\\
Year & 100.00\% & [97.00\%, 100.00\%]\\
Season & 100.00\% & [97.00\%, 100.00\%]\\
State & 99.50\% & [96.50\%, 100.00\%]\\
Operator Affiliation & 98.50\% & [95.50\%, 100.00\%]\\
Agent & 98.50\% & [95.50\%, 100.00\%]\\
Apparatus & 97.00\% & [94.00\%, 100.00\%]\\
Purpose & 99.50\% & [96.50\%, 100.00\%]\\
Target Area & 97.50\% & [94.50\%, 100.00\%]\\
Control Area & 92.00\% & [89.00\%, 95.00\%]\\
Start Date & 98.50\% & [95.50\%, 100.00\%]\\
End Date & 99.50\% & [96.50\%, 100.00\%]\\
\hline
\textbf{Overall Average} & \textbf{98.38\%} & \textbf{[95.38\%, 100.00\%]}\\
\hline
\end{tabularx}
\end{minipage}
\label{tab:field-accuracy-evaluations}
\end{table}

\subsection*{LLM Comparisons}
We evaluated several language models to improve extraction accuracy, 
including \textit{gpt-4.1}, \textit{gpt-4.1-mini}, \textit{gpt-4o-mini}, 
\textit{o4-mini}, and \textit{o3}\cite{openai_models} 
(Table~\ref{tab:model-accuracy-comparison}). To isolate the effect of model choice, we held the prompt constant using a fixed prompt version (Prompt C \cite{llm_extractor_github}) across all evaluations. For each trial, model outputs were compared to human-labeled values across all fields, and the proportion of correct fields was averaged across the dataset.

The \textit{o3} reasoning model achieved the highest overall accuracy at 96.33\%, but was initially an order of magnitude more expensive than \textit{o4-mini} costing approximately \$0.05 per document versus \$0.005 for \textit{o4-mini}, which achieved 95.00\% accuracy. However, in June 2025, OpenAI reduced the price of \textit{o3}, bringing its cost down to \$0.01 per document. Following this price drop, we adopted \textit{o3} to maximize accuracy while maintaining low cost.

\begin{table}[H]
\centering
\caption{Extraction Accuracy by LLM (n=50; prompt C)}
\begin{minipage}{0.85\textwidth}
\begin{tabularx}{\textwidth}{|X|c|c|}
\hline
\textbf{LLM} & \textbf{Accuracy (\%)} & \textbf{95\% Confidence Interval} \\
\hline
gpt-4.1      & 93.33\% & [86.11\%, 97.48\%] \\
gpt-4.1-mini & 93.33\% & [86.11\%, 97.48\%] \\
gpt-4o-mini  & 91.67\% & [84.03\%, 96.34\%] \\
o3           & 96.33\% & [88.08\%, 98.67\%] \\
o4-mini      & 95.00\% & [88.08\%, 98.67\%] \\
\hline
\end{tabularx}
\end{minipage}
\label{tab:model-accuracy-comparison}
\end{table}

\subsection*{Prompt Engineering}
To isolate the effect of prompt design, we measured extraction accuracy with a single model, \textit{o4-mini} (Table~\ref{tab:prompt-accuracy-comparison}). We chose \textit{o4-mini} because it offers a favorable balance between cost and performance. While the absolute accuracies are specific to \textit{o4-mini}, we expect the relative ranking of prompts to extend to other models, such as \textit{o3}.

Three prompt formats were tested: detailed markdown instructions (Prompt A), concise instructions (Prompt B), and a chain-of-thought format (Prompt C) that encouraged step-by-step reasoning. Prompt C achieved the highest accuracy at 95.00\%, suggesting that explicit reasoning steps can substantially improve structured data extraction.

Prompt B performed poorly (44.00\%) due to frequent missing values in the outputs, likely caused by its lack of field-level guidance and examples. In contrast, Prompt A performed better (93.67\%) and benefited from including worked examples for each field. While not explicitly labeled as chain-of-thought, Prompt A was influenced by the same principles, demonstrating that even partial integration of reasoning structure can enhance accuracy.

\begin{table}[H]
\centering
\caption{Accuracy by LLM Prompt (n=50) using \textit{o4-mini}}
\begin{minipage}{0.85\textwidth}
\begin{tabularx}{\textwidth}{|X|c|c|}
\hline
\textbf{Prompt} & \textbf{Accuracy (\%)} & \textbf{95\% Confidence Interval} \\
\hline
Prompt A & 93.67\% & [86.64\%, 97.56\%] \\
Prompt B & 44.00\% & [30.98\%, 57.69\%] \\
Prompt C & 95.00\% & [88.08\%, 98.67\%] \\
\hline
\end{tabularx}
\end{minipage}
\label{tab:prompt-accuracy-comparison}
\end{table}

Together, these validation efforts show that LLM-based data extraction accuracy can be significantly improved through deliberate prompt design and model selection. In particular, prompts incorporating chain-of-thought reasoning and field-specific examples consistently enhanced extraction accuracy, even in complex historical documents with heterogeneous formatting and quality. These findings suggest that careful prompt engineering should be considered a core methodological step when using LLMs to extract scientific data from historical documents.

\section*{Usage Notes}
All weather modification reports in this dataset are self-reported by project operators and may not represent a complete record of U.S. weather modification activity, particularly in cases of unsubmitted or omitted filings. In addition, research field campaigns that piggy-backed on existing cloud seeding operations, such as the ASCII field campaign's monitoring of the Wyoming Weather Modification Pilot Project\cite{geerts2013agi}, may not show up as distinct entries in the data record. 

As of this writing, NOAA’s publicly available records span the years 2000–2025. If older reports (from 1972–1999) are recovered and released, the current extraction pipeline can be applied with minimal modification to extend the dataset’s temporal coverage.

For PDF-to-text conversion, note that \texttt{llm-whisperer} is free only for up to 100 PDF pages per day. Additionally, note that the OpenAI models referenced (e.g., \textit{gpt-4.1}, \textit{gpt-4.1-mini}, \textit{gpt-4o-mini}, \textit{o4-mini}, and \textit{o3}), incur usage costs depending on model and volume, so users should consult current OpenAI pricing for budgeting larger-scale replication \cite{openai_pricing_2025}.

\section*{Data Availability}
The dataset is publicly available on Zenodo (\url{https://doi.org/10.5281/zenodo.14925811}) \cite{final_dataset}. The original NOAA files used as input are publicly available on NOAA (\url{https://library.noaa.gov/weather-climate/weather-modification-project-reports}) \cite{noaa_wm_reports}.

\section*{Code Availability}
Python code for extracting the dataset from the NOAA reports can be obtained
from our public GitHub repository (\url{https://github.com/jdonohue44/NOAA-Weather-Modification-Forms-LLM-Extractor}) \cite{llm_extractor_github}. Python code for visualizing the dataset is also publicly available on GitHub (\url{https://github.com/jdonohue44/US-Cloud-Seeding-Analysis}) \cite{visualize_github}.

\section*{Author Contributions}
All authors have read and agreed to the published version of the manuscript.

\section*{Competing Interests}
The authors declare no competing interests.

\section*{Acknowledgments}
Both authors acknowledge funding from Columbia University's Data Science Institute. K.D.L. acknowledges funding from the Zegar Family Foundation.

\end{document}